\newcommand{\lpx}{\stackrel{\leftarrow}{\partial}_{x}}
\newcommand{\rpx}{\stackrel{\rightarrow}{\partial}_{x}}
\newcommand{\lpp}{\stackrel{\leftarrow}{\partial}_{p}}
\newcommand{\rpp}{\stackrel{\rightarrow}{\partial}_{p}}

\newcommand{\px}{{\partial}_{x}}

\newcommand{\ben}{\begin{equation}}
\newcommand{\een}{\end{equation}}
\newcommand{\bea}{\begin{eqnarray}}
\newcommand{\eea}{\end{eqnarray}}
\newcommand{\nn}{\nonumber\\ }
\newcommand{\dq}{deformation quantization}
\newcommand{\cW}{{\cal W}}
\newcommand{\cWi}{{{\cal W}^{-1}}}

\newcommand{\z}{{\zeta}}
\newcommand{\vp}{{\varphi}}
\renewcommand{\*}{\star}
\renewcommand{\sp}{$\star $-product}
\newcommand{\qq}{\qquad\qquad}
\newcommand{\QQ}{\qquad\qquad\qquad\qquad}
\newcommand{\erf}{{\rm erf}}
\documentclass{article}

\begin{document}

\parskip=4pt
\baselineskip=14pt

%%%%%%%%%%%% Title %%%%%%%%%%

\title{On infinite walls in deformation quantization}
\author{S. Kryukov, M. A. Walton\\\\{\it Department of Physics,
University of Lethbridge}\\
{\em Lethbridge, Alberta, Canada\ \  T1K 3M4}\\
{\small sergei.kryukov@uleth.ca, walton@uleth.ca}\\\\
}

\maketitle
%%%%%%%%%%%%%%%%%%%%%%%%%%%%%%%%%%%%%%%%%%%%%%%%%%%%%%%%%%%%%%%%%%%%%%%%%%%%%%%%%%%%%%%%%%%%%%%%%%%%%%%%%%%%%%%%
\begin{abstract}
We examine the deformation quantization of a single particle
moving in one dimension (i) in the presence of an infinite
potential wall, (ii) confined by an infinite square well, and
(iii) bound by a delta function potential energy. In deformation
quantization, considered as an autonomous formulation of quantum
mechanics, the Wigner function of stationary states must be found
by solving the so-called $\*$-genvalue (``stargenvalue'') equation
for the Hamiltonian. For the cases considered here, this
pseudo-differential equation is difficult to solve directly,
without an ad hoc modification of the potential. Here we treat the
infinite wall as the limit of a solvable exponential potential.
Before the limit is taken, the corresponding $\*$-genvalue
equation involves the Wigner function at momenta translated by
imaginary amounts. We show that it can be converted to a partial
differential equation, however, with a well-defined limit. We
demonstrate that the Wigner functions calculated from the standard
Schr\"odinger wave functions satisfy the resulting new equation.
Finally, we show how our results may be adapted to allow for the
presence of another, non-singular part in the potential.

\end{abstract}

\vfill\eject
%%%%%%%%%%%%%%%%%%%%%%%%%%%%%%%%%%%%%%%%%%%%%%%%%%%%%%%%%%%%%%%%%%%%%%%%%%%%%%%%%%%%%%%%%%%%%%%%%%%%%%%%%%%%%%%%
\section{Introduction}

Deformation quantization\footnote{Deformation quantization is also
known as the Weyl-Wigner(-Moyal) formalism, phase-space
quantization, and by other names as well. For elementary
introductions, see \cite{DQel}. More advanced reviews are listed
in \cite{DQrev}. The modern interpretation of the formalism as a
deformation of classical mechanics was first reported, and
developed extensively, in \cite{BFFLS}.} is touted as a completely
autonomous method of doing quantum mechanics, that is especially
suited to the study of the classical limit.

Some quantum systems that are simple to treat in the Schr\"odinger
formulation, however, are difficult to analyze in deformation
quantization. Even the treatment of a free particle is not
completely straightforward.\footnote{See our Appendix.} Other
examples are a particle moving in one dimension in the presence of
an infinite wall, and a particle confined by an infinite square
well \cite{DP}. Surely, if \dq\ is to take its rightful place as
one of the possible ways of doing quantum mechanics, such simple
systems must be treatable in it.\footnote{MW thanks Brian Wynder
for emphasizing this point to him.}

For stationary, pure states, the $\*$-genvalue (``stargenvalue'')
equation must be solved to find the Wigner function. We will
restrict to such considerations in this work. For the infinite
wall(s) case, simply imposing the usual boundary conditions on its
solutions does not lead to the expected Wigner functions. In
\cite{DP} it was found that if the infinite wall and infinite
square well potentials were modified by an additional ``boundary
potential'' in an ad hoc way, then the expected Weyl transform of
the density operator was a solution. The authors of \cite{DP} did
also show that the added potential terms were consistent with the
Schr{\"o}dinger treatment of those systems. They did not, however,
derive the terms from first principles.

Furthermore, the normal intimate relation of deformation
quantization with classical mechanics is altered by their
modification. In the \dq\ of standard systems, quantum mechanics
is treated as a deformation of classical mechanics, with
deformation parameter $\hbar$. The Hamiltonian is not modified by
quantum corrections, and the deformation is encoded entirely in
the $\*$-product. On the other hand, the ``boundary term'' added
to the potential in \cite{DP} is of order $\hbar^2$, taking the
form $-\frac{\hbar^2}{2m}\delta'(x-a),$ where $x=a$ is the
position of an infinite wall.

Here we study the one-dimensional infinite wall by treating it as
the limit of a solvable exponential potential,
$\lim_{\alpha\rightarrow\infty}\, e^{2\alpha x}$. Our original
hope was to derive the prescription of \cite{DP} from first
principles. We follow instead a different path, however.

Before the limit of the exponential potential is taken, the
$\*$-genvalue equation involves the Wigner function evaluated at
momenta shifted by imaginary amounts. Finding the limit of this
equation directly is problematic. We show, however, that it can be
converted to a partial differential equation, with a well-behaved
limit. The resulting new equation can be treated in the standard
way: no modification of the potential is required, and the normal
Dirichlet boundary conditions can be used. We demonstrate that the
Wigner functions calculated from the standard Schr\"odinger wave
functions satisfy the new equation. Finally, we adapt our results
to allow for the presence of another, non-singular part in the
potential.

We should point out here that even the Schr\"odinger (canonical,
operator) quantization of infinite walls has some subtleties (see
\cite{GK} and references therein). In their operator quantization
of such one-dimensional potentials, the authors of \cite{GK} adopt
an approach similar to ours: they resolve apparent paradoxes by
``acknowledging the existence of the rest of the real line.''
Their treatment has no need for additional boundary potential
terms like those introduced in \cite{DP}.

A quick review of \dq\ is given in section 2. Section 3 is the
meat of the paper, containing the treatments of various cases of
infinite walls and wells described as limits of exponentials. In
section 4, the results are adapted to include an additional,
regular potential, besides those described in section 3. Section 5
is our conclusion, and the Appendix treats the pure \dq\ of a free
particle in one dimension.

%%%%%%%%%%%%%%%%%%%%%%%%%%%%%%%%%%%%%%%%%%%%%%%%%%%%%%%%%%%%%%%%%%%%%%%%%%%%%%%%%%%%%%%%%%%%%%%%%%%%%%%%%%%%%%%
\vskip.5cm\section{Deformation quantization}

This section is a quick review of deformation quantization
\cite{DQel,DQrev}, setting our notation and providing the results
we will need. We will restrict attention to pure states, described
by a state $|\psi\rangle$, or a density operator $\hat\rho =
|\psi\rangle\langle\psi|$.

In deformation quantization, observables are not represented by
operators, but rather by functions on phase space.  They are
multiplied using a pseudo-differential $\*$-product (``star
product'') that is associative but non-commutative. Included is
the Wigner function describing the state of the system, the
central object in deformation quantization. It obeys an evolution
equation involving a $\*$-commutator. In the $\hbar\rightarrow 0$
limit, $\*$-commutators reduce to Poisson brackets, and the
equations of motion of classical mechanics are recovered. In that
sense, this autonomous method describes quantum mechanics as a
deformation of classical mechanics \cite{BFFLS}.

Deformation quantization can be understood as a transform of the
standard way of doing quantum mechanics using the density operator
(matrix). Let us restrict attention (throughout) to the case of a
single particle moving on the $x$-axis, so that phase space has
coordinates $(x,p)$.

In canonical quantization, the phase space coordinates $x$ and $p$
are promoted to the operators $\hat x$ and $\hat p$, obeying the
Heisenberg commutation relation $[\hat x,\hat p]\ =\ i\hbar$.
Observables like $x^2p$ are promoted according to an operator
ordering scheme.\footnote{Deformation quantization may be carried
out in ways that correspond to many different operator orderings.
Here, we will restrict to the Weyl ordering $\cW^{-1}$, resulting
in the famous Gr\"onewold-Moyal $\*$-product. The use of the
inverse notation ($\cW^{-1}$) is for later convenience.} Choosing
the Weyl ordering, our example becomes \ben {\cWi} (x^2p)\ =\
\frac{1}{3}\,(\hat x^2\hat p+\hat x\hat p\hat x + \hat p\hat x^2)\
. \label{Wx2p} \een For a function $f(x,p)$ on phase space, this
generalizes to \ben \cWi f\ =\ f(\partial_a,\partial_b)\, e^{a\hat
x+b\hat p}|_{a,b=0}\ \ ,\label{Wfe}\een where\,
$\partial_a:=\frac{\partial}{\partial a}$, etc. In particular, it
follows that \ben {\cWi}e^{ax+bp}\ =\ e^{a\hat x+b\hat p}\ \
.\label{Wee} \een This gives rise to another expression \ben
{\cWi}f\ =\ \frac{1}{(2 \pi)^2} \int d\tau \, d\sigma \, dx\, dp \
f(x,p) e^{i\tau(\hat p-p) + i\sigma(\hat x-x)}\ ,
\label{Wfint}\een where (\ref{Wee}) has been used in the usual
Fourier formula for $f$.

The crucial property of this ordering is \ben (\cWi f)\,(\cWi g)\
=\ \cWi(f\* g)\ ,\label{cWist}\een proved by Gr\"onewold. The
Gr\"onewold-Moyal \sp\ (pronounced star-product) takes the form
\ben f(x,p)\*g(x,p)\ := \ f(x,p)\, \exp\Big\{\frac{i\hbar}{2}\,
\big(\lpx\rpp-\lpp\rpx\big)\Big\}\,\, g(x,p)\ .\label{Moy}\een
Here the arrows indicate the directions in which the derivatives
act, and they act only on $f$ and $g$, and not to the left or
right of them. That is, eqn. (\ref{Moy}) stands for \ben
f(x,p)\*g(x,p)\ = \ e^{\frac{i\hbar}{2}\,
\big(\partial_{x'}\partial_{p} -
\partial_{p'}\partial_{x}\big)}\,\, f(x',p')
\,g(x,p)\,\vert_{x'=x,\ p'=p}\ .\label{Moyp}\een

Eqn. (\ref{cWist}) shows that the \sp\ is an associative,
non-commutative product of phase space functions that mimics the
product of operators. In deformation quantization, the operator
product is replaced by this \sp, so that operators can be
represented by ordinary functions (and distributions) on phase
space.

The exponent of the \sp\ (\ref{Moy}) indicates the most important
property of \dq: its intimate relation to classical physics. In
classical mechanics, it is the Poisson bracket of functions on
phase space, \ben \{f,g\}\ =\ \frac{\partial f}{\partial x}\,
\frac{\partial g}{\partial p}\ -\ \frac{\partial f}{\partial p}\,
\frac{\partial g}{\partial x}\ =\ f\,(\lpx\rpp-\lpp\rpx)\,g\
\label{Poi}\een that enters the dynamical equations. In the
operator formulation of quantum mechanics, it is the commutator
$[\hat f,\hat g]$ of operator observables $\hat f$ and $\hat g$
that is important. In \dq, the $\*$-commutator \ben  [f,g]_\*\ :=
\ f\* g - g\* f \label{FscG}\een of functions $f$ and $g$ takes
its place. The equation \ben \lim_{\hbar \rightarrow
0}\frac{1}{i\hbar}[f,g]_{\* } = \{ f,g \}\ \ \label{hlim}\een
encodes the correspondence relation between classical and quantum
mechanics in \dq.

The inverse $\cW$ of the Weyl map $\cWi$ is known as the Weyl
transform. One formula for it, \ben {\cW}\hat f\ = \frac{1}{2\pi
\hbar} \, \ \int d\z\, d\vp \,\ Tr(\,\hat f\,\ e^{i[(\hat p-p)\z +
(\hat x-x)\vp ]/\hbar}) \, \ ,\label{cWsym} \een treats momenta
and coordinates in symmetric fashion. A useful variant can be
derived from it: \ben {\cW}\hat f\ =\ \hbar\, \int dy\, e^{-ipy}\,
\langle x+\frac{\hbar y}{2}|\, \hat f\, |x-\frac{\hbar
y}{2}\rangle\ , \label{CTh} \een as well as a similar formula
involving momentum eigenstates. The Weyl transform $\cW\hat f$ of
the operator $\hat f$ is also known as its Weyl symbol.

The object \ben \Delta(x,p\,; \hat x,\hat p)\ :=\ e^{i[(\hat
p-p)\z + (\hat x-x)\vp ]/\hbar}\ ,\label{Strq}\een is prominent in
the fundamental equation (\ref{cWsym}). It is sometimes called the
quantizer, and its importance was stressed in \cite{Str}.

The Weyl transform obeys  \ben \cW(\hat f)\*\cW(\hat g)\ =\ \cW(
\hat f\hat g )\ , \label{cscc}\een an inverse analog of
(\ref{cWist}). This makes it possible to work exclusively with
phase space functions, as long as they are multiplied with the
\sp. Deformation quantization is the realization of this idea - it
is the Weyl transform of quantum mechanics done with the density
operator.

The Weyl transform of the density matrix $\hat\rho=|\psi\rangle\,
\langle\psi|$, \ben \cW\big( \hat\rho\, \big)\ =\ \hbar\, \int
dy\, e^{-ipy}\, \psi^*(x-\frac{\hbar y}{2})\,\psi(x+\frac{\hbar
y}{2})\ , \label{Wig} \een is the central object in deformation
quantization; it describes the quantum state of the system. After
normalization, it is known as the {\it Wigner function}: \ben
\rho\ :=\ \frac{\cW(\hat\rho)}{2 \pi \hbar}\ \ . \label{Wnorm}\een
Combining the last two equations gives \ben \rho[\psi]\ :=\
\frac{1}{2\pi}\, \int dy\, e^{-ipy}\, \psi^*(x-\frac{\hbar
y}{2})\,\psi(x+\frac{\hbar y}{2})\ . \label{Wpsi} \een We will use
the notation $\rho[\psi]$ to emphasize that this is the Wigner
function calculated from a known Schr\"odinger wave function
$\psi$.

The Wigner function evolves according to \ben
{i\hbar}\frac{\partial \rho}{\partial t}\ =\
                [H,\rho]_\*\ .
\label{rhot}\een For stationary states,
$\frac{\partial\rho}{\partial t}=0$, so that \ben [H,\rho]_{\*}\
=\  0\ . \label{Hrcom}\een For $\rho$ describing an energy
eigenstate with eigenvalue $E$,
\begin{equation} H \* \rho\ =\ \rho\* H\ =\  E\rho\ \ .
\label{HrEr}\end{equation}
These simplified dynamical equations allow us to solve for the
Wigner function of the stationary states of the system.

By (\ref{Wig}), the Wigner function can always be calculated if
the Schr\"odinger wave function is known. We are interested here,
however, in considering \dq\ as an autonomous method of doing
quantum mechanics. That means we want to find the Wigner functions
by solving either (\ref{rhot}) or (\ref{HrEr}), and we will
restrict to the latter in this paper. Of course, (\ref{Wig}) can
still be used as a very useful check of our results.

Solving (\ref{HrEr}) can lead to more general solutions than those
of physical interest. Imposing the constraints \ben \rho\*\rho\ =\
\frac{1}{h}\,\rho\ ,\ \ \rho^*\ =\ \rho\ ,\label{rsrs}\een is
important in finding the correct solutions \cite{BFFLS}. Clearly,
the first relation is just the Weyl transform of the usual
projection condition on the density operator. More generally, the
requirement is \ben \rho_i\* \rho_j\ =\
\frac{1}{h}\,\delta_{ij}\,\rho_j\ ,\label{rirj}\een for Wigner
functions describing a discrete set of states, and \ben
\rho_\alpha\* \rho_\beta\ =\
\frac{1}{h}\,\delta(\alpha-\beta)\,\rho_\beta\ \label{rarb}\een
for states with a continuous label.

All observable probabilities can be calculated using the Wigner
function. First of all, by (\ref{Wig}), the probability densities
are \bea |\psi(x)|^2\ =\ \int dp\ \rho(x,p)\ , \qquad\qquad
|\psi(p)|^2\ =\ \int dx\ \rho(x,p)\ . \label{xpden}\eea  Clearly
then, the Wigner function is normalized and real: $\int dx\, dp\,
\rho = 1$ and $\rho^{\ast}=\rho$. The expectation value of an
operator \ben \langle \hat f\rangle\ =\ \int dx\,dp\,\, \rho\* f\
\ , \label{eGints}\een where $f := \cW(\hat f)$. Roughly, one can
think of the integral over phase space as the analog of the trace,
and as discussed above, the star product takes the place of the
operator product. The important cyclic property of a trace is
encoded in \ben \int dx\,dp\,\, f\* g\ =\ \int dx\,dp\,\, g\* f\
=\ \int dx\,dp\,\, f\,g\ .\label{cTr}\een

%%%%%%%%%%%%%%%%%%%%%%%%%%%%%%%%%%%%%%%%%%%%%%%%%%%%%%%%%%%%%%%%%%%%%%%%%%%%%%%%%%%%%%%%%%%%%%%%%%%%%%%%%%%%%%%
\vskip.5cm\section{Free particle except for infinite walls/wells}

In this section, we will treat a particle moving freely on the
$x$-axis, except for the presence of one or more infinite
potential walls or wells. More precisely, the infinite barrier,
the infinite square well, and the delta-function well will all be
studied as limits of potentials built from exponentials.

We are interested in the ``pure'' deformation quantization of
these systems. That is, we would like to derive their Wigner
functions using the equations of deformation quantization only,
without reference to the well known Schrodinger wave functions,
for example. Here we restrict consideration to stationary states.
That means, therefore, that we must examine the $\*$-genvalue
equation (\ref{HrEr}) for the corresponding Hamiltonians.

As mentioned in the introduction, these problems are not
straightforward, as we now explain \cite{DP}. Consider equation
(\ref{HrEr}) more carefully, for the example of an infinite wall,
i.e. for the potential energy \ben V(x)\ =\ \left\{ \matrix{0\, ,
& x<0\, ; \cr \infty\, , & x>0\, .}\right. \een Following the
Schrodinger treatment of this system, one would restrict to $x<0$,
and impose the boundary condition $\rho(0,p)=0$. For $x<0$, the
$\*$-genvalue equation is that of a free particle, with real and
imaginary parts given in the Appendix as equations (\ref{Rfree})
and (\ref{Ifree}), respectively.\footnote{In those equations and
henceforth, we set $\hbar=1$.} But the imaginary part,
$p\,\px\rho\ =\ 0$, does not lead to sensible results for this
potential (see the Appendix).

To look for some guidance, we can study the expected solution, the
Weyl transform of the known density matrix. The Schr\"odinger wave
function is
 \ben \psi(x)=\theta(-x)\left [e^{ i\sqrt E x}
\ -\ e^{ -i\sqrt E x} \right ]\ .\label{psiL}\een Using
 (\ref{Wpsi}), the corresponding
Wigner function is found to be \ben \rho[\psi]\ =\
\theta(-x)\,\bar\rho(x,p)\ ,\ \ \label{rthrb}\een
 with \bea \bar\rho(x,p)\ =\
 &\frac{2\sin[2x(p+\sqrt{E})]}{p+\sqrt{E}}\ +\
 \frac{2\sin[2x(p-\sqrt{E})]}{p-\sqrt{E}} \nn &+\
 2\cos(2x\sqrt{E})\,\frac{2\sin(2xp)}{p}\ .\label{rbwi}\eea
 This $\bar\rho(x,p)$ does not satisfy the imaginary part
(\ref{Ifree}) of the $\*$-genvalue equation $H\*\rho = E\rho$.

Our goal in this article is to find a dynamical equation replacing
the $\*$-genvalue equation that can be solved to find the correct
Wigner functions for these problems.

%%%%%%%%%%%%%%%%%%%%%%%%%%%%%%%%%%%%%%%%%
\subsection{Infinite wall}

The Liouville Hamiltonian is\ben H_\alpha=p^2+e^{2\alpha x}\ .
\label{Halpha} \een  Constants like the mass have been set so as
to simplify considerations ($2m=1$, e.g.). A pure deformation
quantization of this system has already been carried out in
\cite{CFZ}. Taking the $\alpha\rightarrow\infty$ limit of
$V(x)=e^{2\alpha x}$ yields an infinite wall with $V=0$ for $x<0$,
and $V=\infty$ for $x>0$.

Therefore, the limit of the solution found in \cite{CFZ} yields
the correct Wigner function. That is not what we wish to do,
however. We hope instead to take the limit of the $\*$-genvalue
equation, in order to find an equation that can be solved
directly, leading to the physical Wigner function.

The $\*$-genvalue equation is easily found to be \ben
H_\alpha\star \rho(x,p)\ =\ \left[
\left(p-\frac{i}{2}\,\partial_{x}\right)^2 + e^{\, 2\alpha\left(
x+\frac{i}{2}\,\partial_{p}\right)} \right]\rho(x,p)\ =\ E\, \rho
(x,p)\ . \label{Hast} \een It separates into $\Im$ (imaginary)
part \ben \left [-p\,\partial_x+ e^{2\alpha x} \sin (\alpha\,
\partial_{p})\right ]\rho(x,p)\ =\ 0\ ,\label{RHast}\een
and $\Re$ (real) part \ben \left
[p^2-E-\frac{1}{4}\,\partial_{x}^2+ e^{2\alpha x} \cos (\alpha
\,\partial_{p})\right ]\rho(x,p)\ =\ 0\ .\label{IHast}\een
Formally, these equations can be rewritten as  \ben e^{-2\alpha
x}\,\partial_{x}\rho(x,p)\ =-\ \frac{i}{2p}
\left[\rho(x,p+i\alpha)-\rho(x,p-i\alpha)\right ]\ ,
\label{orR}\een and  \bea e^{ -2\alpha x}\left
(p^2-E-\frac{1}{4}\,\partial_{x}^{2}\right )\rho(x,p) \QQ \nn+\
\frac{1}{2}\left [\rho(x,p+i\alpha)+\rho(x,p-i\alpha)\right ]\ =\
0\ .\ \label{orI}\eea

Using (\ref{orR}) to find $\partial_x^2\rho(x,p)$, and
substituting this into (\ref{orI}), we arrive at an equation
without derivatives, \bea 0\ =\ &(p^2-E)\rho(x,p) +
\frac{1}{p}\left (\frac{e^{2\alpha x}}{4}\right )^2
\left[\frac{\rho(x,p+2i\alpha)-\rho(x,p)}{p+i\alpha}
+\frac{\rho(x,p-2i\alpha)-\rho(x,p)}{p-i\alpha}\right ]\nn
 & -\frac{i e^{ 2\alpha
x}}{4p}\left[\rho(x,p+i\alpha)-\rho(x,p-i\alpha)\right]\nn & +
\frac{ e^ {2 \alpha
x}}{2}\left[\rho(x,p+i\alpha)+\rho(x,p-i\alpha)\right]\ .
\label{difff} \eea This can be considered a difference equation in
the momentum variable. Only imaginary shifts of the momentum
arguments are involved; that is, besides $\rho(x,p)$, the result
involves the four quantities \ben \rho(x,p \pm i\alpha)\ ,\ \
\rho(x,p\pm 2i\alpha)\ \ .\label{shifts}\een The
$\alpha\rightarrow\infty$ limit of (\ref{difff}) is problematic.

We can trade the four quantities of (\ref{shifts}), however, for
the derivatives \ben \partial_x^n\rho(x,p)\ ,\ \ n=1,2,3,4 \
.\label{ders}\een The resulting differential equation will have a
well-defined limit as $\alpha\rightarrow\infty$, the result we are
seeking.\footnote{\,H\"older's theorem (see \cite{EDM}, e.g.)
states that no solution of the simple difference equation
$y(x+1)-y(x)=1/x$ satisfies any algebraic differential equation.
Replacing a difference equation with a differential equation may,
therefore, omit interesting solutions. In our case, however, we
will see that the physical Wigner function is a solution of our
resulting differential equation.}

Four equations relating the ``variables'' of (\ref{shifts}) to
those of (\ref{ders}) are required. Two are already provided:
(\ref{orR}) and (\ref{orI}). The two additional equations can be
derived by taking derivatives of (\ref{orI}): \bea &0\ \ =\ \
\partial_{x}^3 \rho(x,p)-4(p^2-E)\,\partial_{x}\rho(x,p) \qq \nn
&+\ 4\alpha e^ {2\alpha x}
\left[\rho(x,p+i\alpha)+\rho(x,p-i\alpha)\right] \nn &-\ 2e^
{2\alpha x}\left[\partial_{x}\rho(x,p+i\alpha)+
\partial_{x}\rho(x,p-i\alpha)\right]\ ,
\label{addi}\eea and \bea 0\ =\
\partial_{x}^{4}\rho(x,p)- 4(p^2-E)\,\partial_{x}^2\rho(x,p)
\qq\qq\qq \nn -\ 8\alpha^2 e^ {2\alpha x} \left[\rho(x,p+i\alpha)+
\rho(x,p-i\alpha)\right] \qq\qq \nn \ \ \ -\ 4 {\alpha}{i} e^
{4\alpha x} \left [ \frac{\rho(x,p+2i\alpha)-\rho(x,p)}{p+i\alpha}
+\frac{\rho(x,p-2i\alpha)-\rho(x,p)}{p-i\alpha}\right] \nn +\ 2 e^
{2\alpha x} \bigg\{\,
\left[4(p+i\alpha)^2-E\right]\rho(x,p+i\alpha) \qq\qq \nn +\ 2 e^
{2 \alpha x}\left[\rho(x,p+2i\alpha)+\rho(x,p)\right] \qq\qq \nn
+\ \left[4(p-i\alpha)^2-E\right] \rho(x,p-i\alpha) \qq\qq \nn +\ 2
e^{ 2 \alpha x} \left[\rho(x,p-2i\alpha)+\rho(x,p)\right]
\,\bigg\}\ .\qq \eea

 With the help of symbolic computation, we find a simple,
 differential equation results:
 \bea 0\ =\ \frac{1}{16}\,\partial_{x}^4 \rho(x,p)+
\frac{1}{2}\,(p^2+E)\,\partial_{x}^2\, \rho(x,p) \qq\qq\nn +\
(p^4-2Ep+E^2)\,\rho(x,p)- e^ {4\alpha x}\,\rho(x,p)\ .\ \ \ \
\label{newde} \eea Taking the limit $\alpha \to \infty $, we find
the new equation  \ben \frac{1}{16}\,\partial_{x}^4 \rho(x,p)+
\frac{1}{2}(p^2+E)\,\partial_{x}^2\, \rho(x,p)+
(p^4-2Ep+E^2)\,\rho(x,p)\ =\ 0\ ,\ \label{newiwall}\een valid for
$x< 0$.

 It is simple to verify that the Wigner function $\bar\rho(x,p)$ of
 (\ref{rbwi}) satisfies the new equation (\ref{newiwall}), for $x<0$.

 It is also interesting to note that eqn. (\ref{newiwall}) can be
 rewritten as \ben \left( p^2\*\rho\* p^2 - E^2\rho\right)\ -\
 2E\,\Re\left(p^2\*\rho-E\rho\right)\ =\ 0\ .\label{HrHetc}\een The new
 equation is a linear combination of ones that follow from the
 $\*$-genvalue equation for a free particle.

%%%%%%%%%%%%%%%%%%%%%%%%%%%%%%%%%%%%%%%%%
\subsection{Infinite square well}

We will now consider the case of two infinite walls, as limits of
two exponential potentials. More precisely, we will be taking the
$\alpha\rightarrow\infty$ limit of the $\*$-genvalue equation
following from the sinh-Gordon \noindent Hamiltonian \ben
H_{\alpha}\ =\ p^2+e^{-2\alpha (x+1)}+e^{2\alpha (x-1)}\ . \een
The method and the result for the two-wall potential will be
remarkably similar to those for the single-wall case.

The $\*$-genvalue equation is  \bea \Bigg\{\left
(p-\frac{i}{2}\,\partial_{x}\right )^2
 + 2 e^{-2\alpha} \cosh\left[2\alpha\left
(x+\frac{i}{2}\,\partial_{p}\right )\right]
 \Bigg\}\, \rho(x,p)\ \nn \ =\ H_{\alpha}\star \rho (x,p)\ =\ E\,
 \rho(x,p)\ .\qq \label{Hawii} \eea It has $\Im$ and $\Re$ parts \ben \partial_{x}\rho(x,p) \ =-\
\frac{ie^{-2\alpha}}{p}
\left[\rho(x,p+i\alpha)-\rho(x,p-i\alpha)\right] \sinh(2\alpha x)\
,\ \label{Risw} \een and \bea \left
(p^2-E-\frac{1}{4}\,\partial_{x}^{2}\right )\rho(x,p)\
\QQ\qq\qq\qq\nn \qq+\ e^{ -2\alpha}\left[\rho(x,p+i\alpha)
+\rho(x,p-i\alpha)\right]\cosh(2\alpha x)\ =\ 0\ .\ \label{Iisw}
\eea

Combining the previous two equations leads to \bea 0\ =\
(p^2-E)\rho(x,p)\ \QQ\QQ\QQ\nn + \frac{e^{-4\alpha}}{4p}
\cosh^2(2\alpha x)\left [\frac{
\rho(x+2i\alpha)-\rho(x,p)}{p+i\alpha} -\frac{
\rho(x,p)-\rho(x,p-2i\alpha))}{p-i\alpha}\right] \nn \ -\
\frac{i\alpha e^{-2\alpha}}{2p} \cosh(2\alpha x)
\left[\rho(x,p+i\alpha)-\rho(x,p-i\alpha)\right] \QQ\nn \ -\ e^{-
2\alpha} \cosh( 2\alpha x)
\left[\rho(x,p+i\alpha)+\rho(x,p-i\alpha)\right]\  .\qq
\label{diffeqsq}\eea This equation involves no derivatives, but
only the quantities \ben \rho(x,p \pm i\alpha),~~~\rho(x,p\pm
2i\alpha) \een as well as $\rho(x,p)$. We again wish to eliminate
these in favor of derivatives $\partial_x^n\rho$, $n=1,2,3,4$. To
do so, we make use of the additional relations \bea
\partial_{x}^3\rho(x,p)\ =\ 4(p^2-E)\,\partial_{x}\rho(x,p)
\QQ\QQ\nn +\ 2\alpha e^{-2\alpha}\sinh( 2\alpha x)
\left[\rho(x,p+i\alpha)+\rho(x,p-i\alpha)\right] \qq\nn -\
4ie^{-4\alpha}\cosh(2\alpha x) \sinh(2 \alpha x)\,\times\, \QQ\nn
\times \left[\frac{ \rho(x,p+2i\alpha)-\rho(x,p) }{p+i\alpha}+
\frac{ \rho(x,p)-\rho(x,p-2i\alpha)}{p-i\alpha}\right]\ ,\
\label{addeqi} \eea and \bea &\,\partial_{x}^{4}\rho(x,p)\ =\
4(p^2-E)\,\partial_{x}^2\rho(x,p)\ \QQ\QQ\nn &+\ 16\alpha^2
e^{-2\alpha} \cosh(2\alpha x)
\left[\rho(x,p+i\alpha)+\rho(x,p-i\alpha)\right] \nn &+\ 16i\alpha
e^{-4\alpha} \sinh^2(2\alpha x)\ \times \QQ\QQ\nn &\times\
\left[\frac{ \rho(x,p+2i\alpha)-\rho(x,p)}{p+i\alpha}+ \frac{
\rho(x,p)-\rho(x,p-2i\alpha) }{p-i\alpha} \right] \nn &+\ e^
{-2\alpha} \cosh(2\alpha x) \bigg\{\left[4(p+i\alpha)^2-E\right]
\rho(x,p+i\alpha) \nn &-\ e^{-2\alpha} \cosh(2\alpha x)
\left[\rho(x,p+2i\alpha)+\rho(x,p)\right] \nn &+\
 \left[4(p-i\alpha)^2-E\right] \rho(x,p-i\alpha) \nn &-\
e^{-2\alpha} \cosh(2\alpha x)
\left[\rho(x,p)+\rho(x,p-2i\alpha)\right]\bigg\}\ .\label{addeqii}
\eea

The resulting new equation is quite complicated, and so we refrain
from writing it here. It does, however, have a well-defined limit
$\alpha \to \infty $, for $x \in\, ]-1,1[$~: \ben
\frac{1}{16}\,\partial_{x}^4 \rho(x,p)+ \frac{1}{2}(p^2+E)
\,\partial_{x}^2 \rho(x,p)+ (p^4-2Ep+E^2)\,\rho(x,p)\ =\ 0\ .\
\label{newiiwall}\een \noindent This is identical to the analogous
result (\ref{newiwall}) found for the one-wall case.

As for the one-wall case, the Wigner functions calculated from the
Schr\"odinger wave functions \ben \psi(x)\ =\
\theta(-x+1)\theta(x+1)\,\cos(\sqrt E x)\ , ~~~ E\ =\
\frac{n^2\pi^2}{4}\ \  \label{pswii}\een satisfy this new
equation, in the following sense. Using (\ref{Wpsi}), we find
$\rho[\psi] = \theta(-x+1)\theta(x+1)\,\bar\rho$, where \bea
\bar\rho(x,p)\ =\
\frac{\sin\left[(2p+n\pi)(1-|x|)\right]}{2p+n\pi}+
\frac{\sin\left[(2p-n\pi)(1-|x|)\right]}{2p-n\pi}\ \nn +\
\frac{\cos(n\pi x)\sin\left[2p(1-|x|)\right]}{p}\ ,\ \qq
\label{Wisw} \eea or, equivalently,  \bea \bar\rho(x,p)\ =\
\frac{\sin\left[2(p+\sqrt{E})(1-|x|)\right]}{2(p+\sqrt{E})}+
\frac{\sin\left[2(p-\sqrt{E})(1-|x|)\right]}{2(p-\sqrt{E})}\ \nn
+\ \frac{\cos(\sqrt{E} x)\sin\left[2p(1-|x|)\right]}{p}\ .\ \qq
\label{Wiswg} \eea $\bar\rho$ satisfies the new equation
(\ref{newiiwall}). As in the one-wall case, the expression for the
Wigner function valid where $V=0$ solves the derived equation.

%%%%%%%%%%%%%%%%%%%%%%%%%%%%%%%%%%%%%%%%%
\subsection{Delta-function potential well}

A $\delta$-function potential well can be studied as the
$\alpha\to\infty$ limit of the Hamiltonian \ben H_{\alpha}=p^2\ -\
2\alpha e^{-2\alpha|x|}\ . \een Restricting to $x>0$, the
$\*$-genvalue equation is \bea H_{\alpha}\star \rho(x,p)\ =\
\left[\left (p-\frac{i}{2}\,\partial_{x}\right )^2 -2\alpha
e^{-2\alpha \left (x+\frac{i}{2}\,\partial_{p}\right
)}\right]\rho(x,p) \nn =\ E \rho(x,p)\ .\QQ \eea Its $\Im$ and
$\Re$ parts are \ben e^{2\alpha x} \,\partial_{x}\rho(x,p)\ =-\
\frac{i\alpha}{p}
\left[\rho(x,p+i\alpha)-\rho(x,p-i\alpha)\right]\ , \een and \bea
e^{2\alpha x}\,\left
(p^2-E-\frac{1}{4}\partial_{x}^{2}\right)\rho(x,p) \QQ\nn -\
\alpha\left[\rho(x,p+i\alpha)\ +\ \rho(x,p-i\alpha))\right]\ =\ 0\
. \eea

Combining these last two relations yields an equation involving no
derivatives, \bea 0=4(p^2-E)\rho(x,p)\ -\ 4\alpha e^{-2\alpha x}
\left[\rho(x+i\alpha)\ +\ \rho(x,p-i\alpha)\right] \qq\nn \ -\
\frac{4\alpha^2 e^{-4\alpha x}}{p} \left[
\frac{\rho(x,p+2i\alpha)-\rho(x,p)}{p+i\alpha} \ +\
\frac{\rho(x,p-2i\alpha)-\rho(x,p)}{p-i\alpha} \right] \nn -\
2i\alpha^2 e^{-2 \alpha x}
\left[\rho(x,p+i\alpha)-\rho(x,p-i\alpha)\right]\ , \qq\eea but
instead depending on $\rho(x,p)$ and \ben \rho(x,p \pm
i\alpha),~~~\rho(x,p\pm 2i\alpha)\ . \een

In the by-now familiar way, we trade the dependence on these for
$x$-derivatives, with the help of the additional equations \bea
\partial_{x}^3 \rho(x,p)\ =\
 4(p^2-E)\,\partial_{x}\rho(x,p) \ \QQ\QQ\nn +\ 8\alpha^2 e^{-2\alpha x}
\left[\rho(x,p+i\alpha)\ +\ \rho(x,p-i\alpha)\right] \QQ\nn +\
4\alpha^2 ie^{-4\alpha x} \left[
\frac{\rho(x,p+2i\alpha)-\rho(x,p)}{p+i\alpha}\ +\
\frac{\rho(x,p)-\rho(x,p-2i\alpha)}{p-i\alpha}\right]\ , \eea and
\bea \partial_{x}^4\rho(x,p)\ =\ 4(p^2-E)\,\partial_{x}^2\rho(x,p)
\QQ\QQ\nn -\ 16\alpha^3 e^{-2\alpha x} \left[\rho(x,p+i\alpha)\ +\
\rho(x,p-i\alpha)\right]\ \QQ\nn  \ -\ 16\alpha^3 e^{-4\alpha x}i
\left[\frac{\rho(x,p+2i\alpha)-\rho(x,p)}{p+i\alpha} \ +\
\frac{\rho(x,p)-\rho(x,p-2i\alpha)}{p-i\alpha} \right]\ \nn -\
4\alpha e^{-2\alpha x} \bigg\{
\left[-4(p+i\alpha)^2+E\right]\rho(x,p+i\alpha) \QQ\nn -\ 4\alpha
e^{-2\alpha x}\left[\rho(x,p+2i\alpha)\ +\ \rho(x,p)\right]+\
\left[-4(p-i\alpha)^2+E\right]\rho(x,p-i\alpha) \ \nn -\ 4\alpha
e^{-2\alpha x} \left[\rho(x,p-2i\alpha)\ +\
\rho(x,p)\right]\bigg\}\ . \qq\eea

The resulting new differential equation \bea
\frac{1}{16}\,\partial_{x}^4 \rho(x,p)\ +\
\frac{(p^2+E)}{2}\,\partial_{x}^2 \rho(x,p)\ \QQ\nn +\
\left[p^4-2Ep+E^2\ +\ 4\alpha^2 e^{-4\alpha x}\right] \rho(x,p)=0\
 \eea has a well-defined limit. Taking $\alpha \to \infty $, gives  \ben
\frac{1}{16}\,\partial_{x}^4 \rho(x,p)\ +\
\frac{1}{2}(p^2+E)\,\partial_{x}^2 \rho(x,p)\ +\
(p^4-2Ep+E^2)\,\rho(x,p)=0\ , \een since $x>0$. The result is
identical for $x<0$, and to the result found above for both the
one-wall and the infinite square well cases.

The sole state bound by the delta-function well has wave function
\ben \psi(x)=e^{-|x|}\ , ~~~ E=-1\ . \een The corresponding Wigner
function $\rho[\psi]$ is \ben \rho(x,p)=\frac{e^{-2x}\left[
\cos(2xp)\ +\ \frac{1}{p}\sin(2xp) \right]}{p^2+1}\ . \een It
satisfies the new differential equation, where $V=0$.

%%%%%%%%%%%%%%%%%%%%%%%%%%%%%%%%%%%%%%%%%%%%%%%%%%%%%%%%%%%%%%%%%%%%%%%%%%%%%%%%%%%%%%%%%%%%%%%%%%%%%%%%%%%%%%%
\vskip.5cm\section{Wall with an additional potential}

So far, we have treated systems that are free except for infinite
walls or wells. Describing such potentials as limits of
exponential terms, we found in all cases that the same
differential equation governs the Wigner functions where $V=0$. It
would be useful to generalize to systems involving additional,
regular potential terms, in the presence of infinite walls/wells.

As a first step, we will consider here the one-wall case, with an
unspecified potential added. That is, we'll consider again the
limit of the Liouville potential, with the additional, regular
potential $V(x)$ present. After a general treatment, we will
consider the simple, special case $V(x)=x^2$ as a check.

The Hamiltonian is \ben H_{\alpha}\ =\ p^2\ +\ e^{2\alpha x} \ +\
V(x)\ , \een yielding the $\*$-genvalue equation \bea
H_{\alpha}\star \rho(x,p)\ =\ E\, \rho (x,p)\ \qq\QQ\QQ\nn =\
\left\{\left (p-\frac{i}{2}\,\partial_{x}\right )^2 +
e^{2\alpha\left (x + \frac{i}{2}\,\partial_{p}\right)}
\right\}\,\rho(x,p)\ +\ V(x)\star \rho(x,p)\ . \eea Its $\Im$ and
$\Re$ parts are \ben \left [-p\,\partial_x\ +\ e^{2\alpha x} \sin(
\alpha \,\partial_{p})\right ]\,\rho(x,p)\ +\ \Im\left[V(x)\star
\rho(x,p)\right]\ =\ 0\ , \een and \ben \left[
p^2-E-\frac{1}{4}\,\partial_{x}^2\ +\ e^{2\alpha x} \cos(\alpha
\,\partial_{p})\right]\rho(x,p)\ +\ \Re\left[V(x)\star
\rho(x,p)\right]\ =\ 0\ . \een Combining these gives  \bea 0\ =\
4(p^2-E)\,\rho(x,p)\ \qq\QQ\QQ \nn +\ \frac{e^{4\alpha x}}{4p}
\left[\frac{\rho(x\ +\ 2i\alpha)\ -\ \rho(x,p)}{p+i\alpha}\ +\
\frac{\rho(x-2i\alpha)\ -\ \rho(x,p)}{p-i\alpha}\right]\  \nn \ +\
\frac{i\alpha e^{2\alpha x}}{p} \left[\rho(x,p+i\alpha)\ -\
\rho(x,p-i\alpha)\right]\ \QQ\nn +\  2 e^ {2 \alpha
x}\left[\rho(x,p+i\alpha)\ +\ \rho(x,p-i\alpha)\right]\ \QQ\nn  +\
4\Re\left[V(x)\star \rho(x,p)\right]\ -\ \frac{1}{p}\,\partial_{x}
\Im\left[V(x)\star \rho(x,p)\right] \QQ\nn \ +\ \frac{ie^{2\alpha
x}}{2p} \left\{
\frac{\Im\left[V(x)\star\rho(x,p+i\alpha)\right]}{p+i\alpha} \ -\
\frac{\Im
\left[V(x)\star\rho(p-i\alpha)\right]}{p-i\alpha}\right\}\ . \eea

Following the procedure used above, we would like to eliminate the
quantities $\rho(x,p \pm i\alpha),~\rho(x,p\pm 2i\alpha)$ in favor
of derivatives of the Wigner function. As in the cases previously
considered, two additional equations can be derived
straightforwardly:  \bea 0\ =\ \ -\
\partial_{x}^3 \rho(x,p)\ +\ 4(p^2-E)\,\partial_{x}\rho(x,p)
\qq\QQ\nn +\ 4\alpha e^{2\alpha x} \left[\rho(x,p+i\alpha)\ +\
\rho(x,p-i\alpha)\right]\  \QQ \nn \ -\ ie^{4\alpha x}
 \left[
\frac{\rho(x,p+2i\alpha)\ -\ \rho(x,p)}{p+i\alpha}\ +\
\frac{\rho(x,p)\ -\ \rho(x,p-2i\alpha)  }{p-i\alpha} \right] \nn \
+\ 2e^{2\alpha x} \left[ \frac{\Im\left[V(x)\star \rho (x,p+ i
\alpha)\right]}{p+i \alpha} \ +\ \frac{\Im\left[V(x)\star \rho
(x,p-i \alpha)\right]}{p-i \alpha} \right]\  \nn \ +\
4\,\partial_{x}\Re\left[V(x)\star \rho(x,p)\right]\ , \qq \eea and
\bea 0\ =\ \ -\
\partial_{x}^{4}\rho(x,p) \ +\ 4(p^2-E)\partial_{x}^{2}\rho(x,p)
\QQ\QQ \nn +\ 8\alpha^2
e^{2\alpha x} \left[\rho(x,p+i\alpha)\ +\
\rho(x,p-i\alpha)\right] \QQ\nn \ -\ 4 i\alpha e^{4\alpha x}
\left[
 \frac{\rho(x,p+2i\alpha)\ -\ \rho(x,p)}{p+i\alpha} \ +\
\frac{\rho(x,p)\ -\ \rho(x,p-2i\alpha)}{p-i\alpha}\right ] \nn \
+\ 2e^{2\alpha x} \bigg\{\,
\left[4(p+i\alpha)^2-E\right]\rho(x,p+i\alpha) \QQ\nn  +\ 2 e^{2
\alpha x}\left[\rho(x,p+2i\alpha)\ +\ \rho(x,p)\right] \qq\qq\nn \
+\ \left[4(p-i\alpha)^2-E)\right] \rho(x,p-i\alpha)\ +\  2 e^{2
\alpha x} \left[\rho(x,p-2i\alpha)\ +\ \rho(x,p)\right] \,\bigg\}
\nn \ +\ 8\alpha e^{2\alpha x} \left\{ \frac{\Im\left[V\star
\rho(x,p+i\alpha)\right]}{p+i\alpha} \ +\ \frac{\Im\left[V\star
\rho(x,p-i\alpha)\right]}{p-i\alpha} \right\}  \nn +\ 8 e^{2\alpha
x} \left\{\Re\left[V (x)\star\rho(x,p+i\alpha)\right]\ +\
\Re\left[V(x)\star\rho(x,p-i\alpha)\right]\right\} \nn +\
4\,\partial_{x}^2\Re\left[V(x)\star \rho(x,p)\right]\ . \QQ\qq
\eea This is not sufficient in this more general case, however. We
now have 4 further quantities \ben \left ({\Re}/{\Im}\right
)\Bigg[V(x)\star \rho(x,p\pm i\alpha)\Bigg]\ , \een to get rid of.
Here $\left ({\Re}/{\Im}\right )$ indicates the real part or the
imaginary part.

With considerably more work, however, the required 4 equations can
be derived. We find \bea \left ({\Re}/{\Im}\right )
\Bigg[V(x)\star \left(
p^2-E-\frac{1}{4}\,\partial_{x}^2\rho(x,p)\right) \Bigg]\ \qq\QQ
\nn \ =\ \ -\ \frac{e^{2 \alpha x}}{2}\left ({\Re}/{\Im} \right )
\Bigg[V(x)\star \Big[\rho(x,p+i\alpha)\ +\
\rho(x,p-i\alpha)\Big]\Bigg] \nn -\ \left ({\Re}/{\Im}\right
)\Bigg[V(x)\star \Re \Big[V(x)\star\rho(x,p)\Big]\Bigg]\ , \QQ
\eea and \bea -\left ({\Re}/{\Im}\right ) \Bigg[V(x)\star p\,
\partial_{x}\rho(x,p)\Bigg] \QQ\QQ \nn -\ \frac{ie^{2 \alpha x}}{2} \left
({\Re}/{\Im}\right )\left\{V(x)\star \left[\rho(x,p+i\alpha)\ +\
\rho(x,p-i\alpha)\right]\right\}\  \quad\nn \ +\ \left
({\Re}/{\Im}\right )\left\{V(x)\star \Im
\Big[V(x)\star\rho(x,p)\Big]\right\}\ =\ 0\ . \qq\eea

By symbolic computation, the resulting differential equation can
be derived. In the limit, it goes to  \bea
\frac{1}{16}\,\partial_{x}^4\, \rho(x,p)\ +\
\frac{(p^2+E)}{2}\,\partial_{x}^2\, \rho(x,p)\ +\
(p^4-2Ep+E^2)\rho(x,p)\ \nn \ +\ (p^2-E) \Re
\left[V(x)\star\rho(x,p)\right] \ -\ p\,\partial_{x}
\Im\left[V(x)\star \rho(x,p)\right] \qq \nn -\
\frac{1}{4}\,\partial_{x}^2 \Re \left [V(x)\star\rho(x,p)\right]\
-\ \Im \left[V(x)\star{p}\,\partial_{x}\rho(x,p)\right] \qq \nn +\
\Im \left\{ V(x)\star \Im \left[V(x)\star\rho(x,p)\right] \right\}
\ +\ \Re\left\{V(x)\star \Re \left[V(x)\star\rho(x,p)\right]
\right\}\ \nn \ +\ \Re\left\{V(x)\star \left [\left
(p^2-E-\frac{1}{4}\,\partial_{x}^2\right )\rho(x,p)\right
]\right\}\ =\ 0\ , \label{showeqn}\eea for $x< 0$. Clearly, this
equation reduces to the one found above when $V=0$.

As a simple check of this result, consider the example of the
simple harmonic potential $V(x)=x^2$. Only the odd-parity wave
functions have $\psi(0)=0$, and so only they survive the presence
of the infinite potential wall at $x=0$. The ground state wave
function is therefore \ben \psi(x)\ =\ \theta(-x)\,\,x\, e^{-x^2
}\ ,\een with energy $E=3$. The corresponding Wigner function is
$\rho[\psi]=\theta(-x)\bar\rho$ with \bea \bar\rho(x,p)\ =\  x^2
\,\erf(x-ip)\,\pi e^{-p^2-x^2}\ \qq\QQ \nn \ -\ \frac{1}{2}
\,\erf(x+ip)\pi \,e^{-p^2-x^2} \ +\ x^2 \,\erf(x+ip)\pi
e^{-p^2-x^2}\ \nn  \ +\ \sqrt \pi x\,e^{-2x(x-ip)} \ +\ i\sqrt \pi
p \,e^ {-2x(x-ip)} \qq \nn +\ \erf(x-ip)p^2 \pi e^ {-p^2-x^2}\ -\
\frac{1}{2} \,\erf(x+ip)\pi e^{-p^2-x^2}
 \nn +\ \sqrt \pi x e^{-2x(x+ip)}
\ -\ i\sqrt \pi p e^{-2x(x+ip)}\ \nn +\ \erf(x-ip)p^2 \pi
e^{-p^2-x^2}\ . \QQ \label{rshowall}\eea Here $\erf$ denotes the
error function \ben \erf(x)\ =\ \int_{0}^{x} e^{-t^2}\, dt\ . \een
We have verified that (\ref{rshowall}) does indeed satisfy the
equation (\ref{showeqn}).

%%%%%%%%%%%%%%%%%%%%%%%%%%%%%%%%%%%%%%%%%%%%%%%%%%%%%%%%%%%%%%%%%%%%%%%%%%%%%%%%%%%%%%%%%%%%%%%%%%%%%%%%%%%%%%%
\vskip.5cm\section{Conclusion}

If deformation quantization is truly an autonomous formulation of
quantum mechanics, problems that are solved simply using other
methods must be treatable in it. First-principle \dq\ of the
systems should be possible, even if difficult. For stationary
states, the Wigner functions should be derivable by solving the
$\*$-genvalue equation for the Hamiltonian.

Here we have studied a single quantum particle travelling freely
on the $x$-axis, except for the presence of one or two infinite
walls. The simple Schr\"odinger wave functions of these systems
are well known. The Wigner functions expected from the
stationary-state wave functions can therefore be easily worked out
(see $\rho[\psi]$ in eqn (\ref{Wpsi})). Dias and Prata \cite{DP}
point out, however, that they do not satisfy the corresponding
$\*$-genvalue equations.

To cure this problem, they propose the addition of ``boundary
potentials'' to the Hamiltonian. They further show that these
additional terms are consistent with the Schr\"odinger
quantization of the systems. They do not, however, derive the
extra potentials from first principles.

In the hopes of filling this gap, we treated the infinite wall and
infinite square well potentials as limits of solvable, exponential
ones. The delta-function well is also considered, since it can be
treated in a very similar way. Our results do not relate easily to
the proposal of \cite{DP}, however.

A well-defined limit was found for the corresponding $\*$-genvalue
equations. Happily, this new differential equation is common to
all the systems. Furthermore, the expected Wigner functions
$\rho[\psi]$ satisfy it away from the infinite walls. The revised
hope is therefore that the new equation can be solved to find the
correct Wigner functions directly, and boundary conditions on the
Wigner functions can be imposed in the way familiar from
Schr\"odinger quantization.

That hope has not been realized here. We have not (yet) shown that
the new differential equations can be solved to find the correct,
physical Wigner functions. We only verified that the
$\rho[\psi]$'s satisfied the new equations, outside the $V=\infty$
regions.

Nevertheless, finding a well-defined limit of the $\*$-genvalue
equations for these systems is a significant first step. The new
equation was also generalized to allow for a regular potential in
addition to infinite walls/wells.

It might be interesting to use other solvable potentials instead
of the exponential ones. It would at least (hopefully) verify that
our results are independent of our particular choice of
representation of the infinite potential wall as a limit. For
example, $\lim_{N\rightarrow\infty}x^N$ yields the infinite square
well. Aspects of this limit have already been studied in
\cite{Vo}.

\vskip1cm
%%%%%%%%%%%%
\appendix{\large{\bf Appendix:\ \  Free particle}} \vskip.5cm

In the reviews we have seen, a pure deformation quantization of a
free particle moving in one dimension is conspicuously absent.
Here we attempt such a treatment. If nothing else, this appendix
should indicate why the free example is typically omitted.

Putting $2m=1$ to get the Hamiltonian $H=p^2$, the $\*$-genvalue
equation becomes \ben \bigg(p-\frac{i\hbar}{2}\px\bigg)^2\rho\ =\
E\rho\ .\label{free}\een With $\hbar=1$, the imaginary part of
this equation is \ben p\,\px\rho\ =\ 0\ ,\label{Ifree}\een while
\ben \bigg(p^2-\frac{1}{4}\px^2\bigg)\rho\ =\ E\rho\
.\label{Rfree}\een is the real part.

The factor $p$ in (\ref{Ifree}) is crucial: we can conclude that
when $p\not=0$, $\px\rho=0$, but not when $p=0$. Substituting the
ansatz \ben \rho(x,p)\ =\ f(p) + \delta(p)\,g(x)\ ,\label{ans}\een
in (\ref{Rfree}), we find \ben (p^2-E)f(p)\ -\ \delta(p)\bigg( E+
\frac{1}{4}\px^2\bigg)g(x)\ =\ 0\ .\label{Rfi}\een Considering
$p=0$ yields \ben g(x)\ =\ b\exp\bigg(2i\sqrt{E}x\bigg)\ +\
b^*\exp\bigg(-2i\sqrt{E}x\bigg)\ , \label{bpm}\een where
$\rho=\rho^*$ has been imposed. Then (\ref{Rfi}) reduces to
$(p^2-E)f(p)=0$, solved by \ben f(p)\ =\ a_+\delta(p-\sqrt{E})\ +\
a_-\delta(p+\sqrt{E})\ ,\label{apm}\een with $a_\pm$ arbitrary
real constants.

The terms of (\ref{apm}) correspond to plane waves of momentum
$\pm\sqrt{E}$, as can be verified by solving \ben p\*\rho\ =\
\rho\* p\ =\ \sqrt{E}\rho\ .\label{eigp}\een The expression \ben
\rho\ =\ a_+\delta(p-\sqrt{E})\ +\ a_-\delta(p+\sqrt{E})\
,\label{rapm}\een  is the Wigner function of a mixed state of the
two momentum eigenstates. The terms of (\ref{bpm}) are necessary
for coherent superpositions of the two momentum eigenstates, and
they represent interference between them. That they are required
can be seen by considering the Wigner function of a simple
harmonic oscillator in the long-period limit, or the large-width
limit of a particle in an infinite square well.

The general result is  \bea \rho\ =\ &\delta(p)\,\bigg\{\,
b\exp\big(2i\sqrt{E}x\big)\ +\
b^*\exp\big(-2i\sqrt{E}x\big)\,\bigg\}\ \nn &+\ \ \
a_+\delta(p-\sqrt{E})\ \ +\ \ a_-\delta(p+\sqrt{E})\ \
.\label{rhog}\eea

To restrict to pure-state Wigner functions, we impose \ben
\rho\*\rho\ \propto\ \delta(0)\,\rho\ , \label{pure}\een valid for
the Wigner functions corresponding to non-normalizable pure
states. We find \bea \rho\*\rho\ &=\ \delta(0)\, \bigg\{\,
(a_+^2+|b|^2)\delta(p-\sqrt{E})\ +\
(a_-^2+|b|^2)\delta(p+\sqrt{E})\ \nn &+\ (a_++a_-)\delta(p)\,\big[
b\exp(2i\sqrt{E}x)\ +\ b^*\exp(-2i\sqrt{E}x) \big]\, \bigg\}\
.\label{rr}\eea The calculation is straightforward, except that we
interpret $\delta(p+\sqrt{E})\delta(p+\sqrt{E})$ as
$\delta(0)\delta(p+\sqrt{E})$, e.g., since they yield equivalent
results as distributions (i.e. upon integration). Similarly,
assuming $E>0$, we put $\delta(p+\sqrt{E})\delta(p-\sqrt{E})$ to
zero, e.g.

In (\ref{pure}), this result yields the constraint \ben |b|^2\ =\
a_+a_-\ \ \Rightarrow\ \ b\ =\ \sqrt{a_+a_-}\, e^{i\phi}\ ,\
\phi\in{\Re}\, . \label{purecon}\een The general pure-state
solution to the free-particle $\*$-genvalue equation (\ref{free})
is therefore \bea \rho\ =\ &a_+\delta(p-\sqrt{E})\ \ +\ \
a_-\delta(p+\sqrt{E})\ \ \ \nn &\ \ +\ \
2\sqrt{a_+a_-}\,\,\delta(p)\,\cos\big(2\sqrt{E}x+\phi\big)\  \
.\label{rhopure}\eea

On the other hand, calculating \ben \rho[\psi]\ :=\
\frac{1}{2\pi}\,\int dy\,
e^{-ipy}\psi^*(x-\frac{y}{2})\psi(x+\frac{y}{2})\
,\label{rhop}\een with the pure-state wave function \ben \psi\ =\
\alpha_+e^{i\sqrt{E}x}\ +\ \alpha_-e^{-i\sqrt{E}x}
\label{psif}\een yields \bea \rho[\psi]\  =&\ |\alpha_+|^2
\delta(p-\sqrt{E})\ +\ |\alpha_-|^2 \delta(p+\sqrt{E}) \nn &+\
\delta(p)\bigg\{ \alpha_+^*\alpha_-e^{-2i\sqrt{E}x}\ +\
\alpha_+\alpha_-^*e^{2i\sqrt{E}x} \bigg\}\ .\label{rprp}\eea
Comparing (\ref{rprp}) with (\ref{rhopure}) reveals a one-to-one
correspondence, however, given by the relations \ben \alpha_\pm\,
 =\, \sqrt{a_\pm}\,\,e^{i\phi_\pm}\ \ ,\ \ \ \phi-\phi_++\phi_-\,=\, 0\
.\label{preln}\een As should be, only the relative phase
$\phi_+-\phi_-$ of (\ref{psif}) is relevant to the Wigner
function.

\vskip1cm \noindent{\bf Acknowledgments}\hfill\break We thank
Terry Gannon, Tom Osborn and J{\o }rgen Rasmussen for helpful
comments on the manuscript. This research was supported in part by
NSERC of Canada, and by the University of Lethbridge. SK is also
grateful for a NATO Science Fellowship.

%%%%%%%%%%%%%%%%%%%%%%%%%%%%%%%%%%%%%%%%%%%%%%%%%%%%%%%%%%%%%%%%%%%%%%%%%%%%%%%%%%%%%%%%%%%%

\end{document}